\documentclass[journal]{IEEEtran}

\usepackage[T1]{fontenc}
\usepackage{amsmath,amsfonts,amssymb,bm,mathrsfs}
\usepackage{graphicx,color}
\usepackage{subcaption}
\usepackage[dvipsnames]{xcolor}
\usepackage[colorlinks=true,linkcolor=black,citecolor=black,urlcolor=black]{hyperref}
\usepackage{url}
\usepackage{cite}

\newcommand{\LRHeader}{D. Teyssieux, J Millo, E. Rubiola, R Boudot \emph{Phase noise of a microwave photonic channel: direct-current versus external electro-optic modulation}.}
\title{Phase noise of a microwave photonic channel: direct-current versus external electro-optic modulation
  \author{%
    Damien Teyssieux, Jacques Millo, Enrico Rubiola, and Rodolphe Boudot
    }
  \thanks{%
    This work is partially funded by the Agence Nationale de Recherche (ANR) Programme d'Inves\-tis\-se\-ment d'Avenir under the following grants:
    ANR-11-EQPX-0033-OSC-IMP (Oscillator IMP project)
    ANR-10-LABX-48-01 (FIRST-TF network), and
    ANR-17-EURE-00002 (EIPHI); 
    and by grants from the Région Bourgogne Franche Comté intended to support the above.
  }
  \thanks{
    The authors are with 
    Université de Franche Comté (uFC), 
    Centre National de la Recherche Scientifique (CNRS),
    Institut FEMTO-ST,  
    and Supmicrotech-ENSMM,
    Besancon, France.
    }
  \thanks{Enrico Rubiola is also with the Divsion of Quantum Metrology and Nanotechnology, Istituto Nazionale di Ricerca Metrologica (INRiM), Torino, Italy.}
  \thanks{\emph{Corresponding author}: R. Boudot, \texttt{rodolphe.boudot@femto-st.fr}}
}

\begin{document}
\maketitle

\begin{abstract}
\boldmath
We characterize the phase noise of a microwave photonic channel, where a 10 GHz signal is carried by an intensity-modulated light beam over a short optical fiber, and detected.  Two options are compared, (i) an electro-optic modulator (EOM), and (ii) the direct modulation of the laser current.  The 1.55~$\mu$m laser and the detector are the same.  The effect of experimental parameters is investigated, the main of which are the microwave power and the laser bias current. The main result is that the upper bound of the phase flicker is $-117$~dBrad$^2$ in the case of the EOM, limited by the background noise of the setup.  In contrast, with direct modulation of the laser, the flicker is of $-114$ to $-100$~dBrad$^2$, depending on the laser bias current (50--90~mA), and the highest noise occurs at the lowest bias.
Our results are of interest in communications, radar systems, instrumentation and metrology.
\unboldmath
\end{abstract}

\section{Introduction}\label{introduction}
The generation of low-phase-noise microwaves from optics has found great interest and extensive use in applications such as high-performance Doppler radar systems \cite{Ghelfi:Nature:2014}, communications \cite{Koenig:2013}, low-timing jitter analog-digital conversion \cite{Valley}, and time and frequency metrology \cite{Nakamura:2020, Nardelli:2022}.  Additionally, low noise microwaves carried by an optical beam are required in vapor cell atomic clocks based on coherent population trapping (CPT) \cite{MAH:APL:2018} because the phase noise of the microwave field which interrogates the atoms can limit the clock's short-term stability \cite{Danet:UFFC:2014}.

The purest microwaves are nowadays obtained by frequency-division from cavity-stabilized lasers using optical frequency combs (OFCs) \cite{Fortier:2011}.  The phase noise of such systems is of the order of $-$172 and $-$107 dBrad$^2$/Hz at 100~kHz and 1~Hz Fourier frequency, respectively, compliant with zeptosecond-level time fluctuations \cite{Xie:2016}. However, ultra-stable optical cavities and OFCs are rather large and fragile pieces of equipment, compared to regular oscillators and synthesizers. The cavity requires extreme mechanical and thermal stability, while the OFC relies on the generation of octave-wide supercontinuumm light and on the simultaneous stabilization of repetition rate and carrier-envelope frequency offset.  Therefore, moving such systems outside metrological labs is a challenge.  Simpler and compact systems have been proposed \cite{Giunta:2020}, relying on the use of a free-running monolithic femtosecond laser \cite{Kalu:2020}, or on the transfer oscillator technique \cite{Telle:2002, Nardelli:2022}, possibly coupled with soliton-microcombs \cite{Gaeta:2020, Liu:2020}.

The optoelectronic oscillator (OEO) is an alternative option for the generation of low noise microwaves \cite{Yao:1996, Maleki:2011, Chembo:RMP:2019, Tuto:2020}.  The OEO is a delay-line oscillator, where the microwave delay is implemented with a photonic channel (laser, intensity modulator, optical fiber and photodetector).  The low loss of the fiber (0.2~dB/km at 1.55~$\mu$m wavelength) enables the implementation of a long delay, which is equivalent to a large-$Q$ resonator in the feedback loop.  For example, a delay of 10 $\mu$s (2 km fiber) is equivalent to $Q=~3.14{\times}10^5$ at 10~GHz carrier, out of reach for room-temperature resonators (Ref.~\cite[Chap.~3-4]{Rubiola:oscillators}).

\begin{figure*}[t]   
{\centering\includegraphics[width=\linewidth]{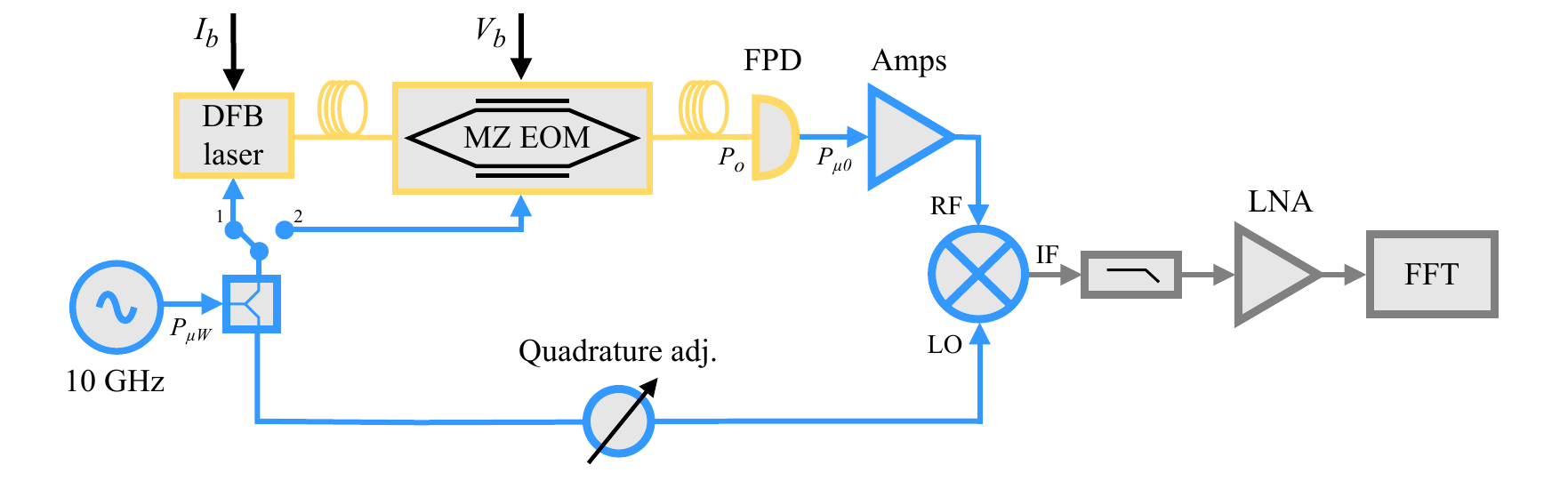}}
\caption{Simplified schematic of the experimental setup used for phase noise measurements at 10 GHz of the microwave-modulated laser system.  The 10 GHz microwave reference is split into two arms.  Setting the switch in position `1' (actually, moving a semirigid cable and terminating the unused input), the microwave modulates the laser power via the bias current, and the MZ EOM is replaced with a fiber patch.  In position `2,' the laser is in CW mode, and the beam is modulated by the EOM\@.  The fast photodiode (FPD) extracts the microwave, which is amplified and compared to the reference.  With the RF and LO inputs saturated and in quadrature, the mixer works as a phase-to-voltage converter.  The Fast Fourier Transform (FFT) analyzer measures the power spectral density averaged over a convenient number of acquisitions, which smooths the phase-noise plots.}
\label{figure1}
\end{figure*}

Two main modulation configurations are seen in OEOs and in CPT clocks, namely, (i) a fibered Mach-Zehnder electro-optic modulator (EOM) at the output of a CW laser diode, and (ii) the direct modulation (DM) of a laser diode.  The former option is most often used, and is encountered in the demonstration of state-of-the-art OEOs \cite{Tuto:2020} and CPT-based atomic clocks \cite{MAH:APL:2018}.  That said, there are good reasons to abandon the EOM in favor of the direct modulation.  The EOM is large ($\approx10$~cm long) and expensive, it has large optical loss (4--5 dB, plus 3 dB intrinsic loss due to the idle point at half power), shows large temperature sensitivity because of the $\mathrm{LiNbO_3}$ waveguide, and requires large microwave power ($\approx50{-}100$~mW).  In turn, this is detrimental to the thermal stability, and stabilization of the bias point may be necessary \cite{Liu:PRA:2013}.
Conversely, the direct modulation of the laser requires only an appropriate network to combine DC bias and microwave in the laser diode.
Several OEO structures based on directly modulated distributed-feedback lasers \cite{Wishon:PTL:2018, Qi:2021, Sinquin:2021}, vertical-cavity surface-emitting lasers \cite{Hasegawa:PTL:2007} and microsquare \cite{Liao:2018} lasers have been demonstrated, with phase noise levels reaching $-129~\mathrm{dBrad^2/Hz}$ at 10~kHz Fourier frequency for a 10~GHz carrier.  With the same motivation, high-performance CPT-based atomic clocks using directly-modulated lasers have been reported, exhibiting competitive short-term stability \cite{Yun:Metrologia:2021}.

The studies discussed focus on the characterization of the overall phase noise of the output microwave signal, with comparatively little attention to the noise contribution of the photonic channel, i.e., microwave-to-microwave via modulated light.  Yet, the quantitative knowledge of such contribution is necessary to assess ultimate phase noise limit.  The phase noise of optical links using DM lasers was considered in \cite{Bibey:MTT:1999}, but reporting a limited number of experimental cases.  Other studies focus on the numerical \cite{Ahmed:IEEE:2001} or electrical modelling \cite{Bdeoui:EMC:2003, Kassa:IEEE:2013} of DM laser systems. A theoretical analysis on the noise of links using externally-modulated lasers is reported in \cite{Qi:JLT:2006}. 
Reference \cite{Teyssieux:JOSAB:2022} demonstrates a technique to mitigate the microwave phase noise of a DM laser using feedback on the laser current.  Reference \cite{Boudot:APL:2022} reports on a self-sustained microwave oscillator, based on a EOM and on a CPT cesium cell as the resonator in the feedback loop,  with a phase noise in agreement with the Leeson model \cite{Leeson:1966, Rubiola:oscillators}. 

In this article, we compare the phase noise of a 10 GHz photonic channel in the two relevant configurations mentioned, CW laser followed by an EOM, and directly modulated laser.  The flicker phase noise is lower than $-117$ dBrad$^2$ in the case of the EOM-based setup.  This is an upper bound, limited by the background of the measurement system.  For comparison, the flicker of high-speed photodetectors similar to ours is of $-120$ dBrad$^2$ or lower \cite{Rubiola:PDs:2006}.  We found no data about the flicker of the EOM, but our experience suggests that it is negligible at this scale.  In contrast, the direct modulation gives a flicker of $-114$~dBrad$^2$ at 90 mA bias current, progressively increasing to $-100$~dBrad$^2$ when the bias current is reduced to 50 mA\@.  The higher noise, compared to the CW laser plus EOM, is clearly due to the modulation of the laser bias current.

\section{Experimental setup}
Figure \ref{figure1} shows a schematic of the experimental setup. The reference microwave source is a Rohde \& Schwarz SMB100A\@.  The laser source is a single-mode pigtailed distributed feedback (DFB) diode laser (Gooch and Housego AA0701) with internal isolator, emitting at 1550 nm wavelength.  The laser is driven by a low noise current controller \cite{Libbrecht:RSI:1993} and is integrated in a Butterfly package with embedded thermistor and Peltier cooler.  The laser has an efficiency of $\sim 0.19$~W/A beyond the 10~mA threshold, delivering 16 mW optical power at 90 mA bias current.
The laser is followed by a voltage-controlled optical attenuator (VOA, IDIL COCOM03898, not shown on Fig.~1), which is used to set the same optical power at the photodetector input for the two configurations.
The EOM (iXblue MXAN-LN-20, $V_{\pi}$~$\simeq$~5.5~V) has no temperature control, and no active control of the bias voltage $V_b$. It has a polarization controller at the input. 
The fast photodiode, a Discovery DSC30, is followed by four cascaded HMC606 amplifiers with 9 dB attenuation between $2^\mathrm{nd}$ and $3^\mathrm{rd}$, providing a total gain of 35 dB and a noise figure of 4.8 dB.  The output power at the amplifier output is measured with a power meter (Rohde-Schwarz NRVS) tapping the signal with a 10-dB directional coupler (Macom PN2020).  The mixer (Miteq DB0218) is followed by a 1.9 MHz lowpass filter (Mini-Circuits SLP-1.9+) which eliminates the 2nd harmonics (20 GHz) and unnecessary high-frequency noise, followed by a 40 dB low-noise amplifier optimized for lowest flicker \cite{Rubiola:RSI:2004}.
The fast Fourier transform (FFT) analyzer is a HP3562A\@.  However old, it features efficient logarithmic frequency resolution.
A line stretcher (Arra 9426B) in the lower arm of the system enables fine adjustment of the quadrature relation at the mixer inputs.
The background noise of the setup is measured by replacing the photonic channel with a variable attenuator (Arra 6803-10A), so that the microwave power remains the same as in the two configurations under test.  


\begin{figure}[t]
\centering
\includegraphics[width=\linewidth]{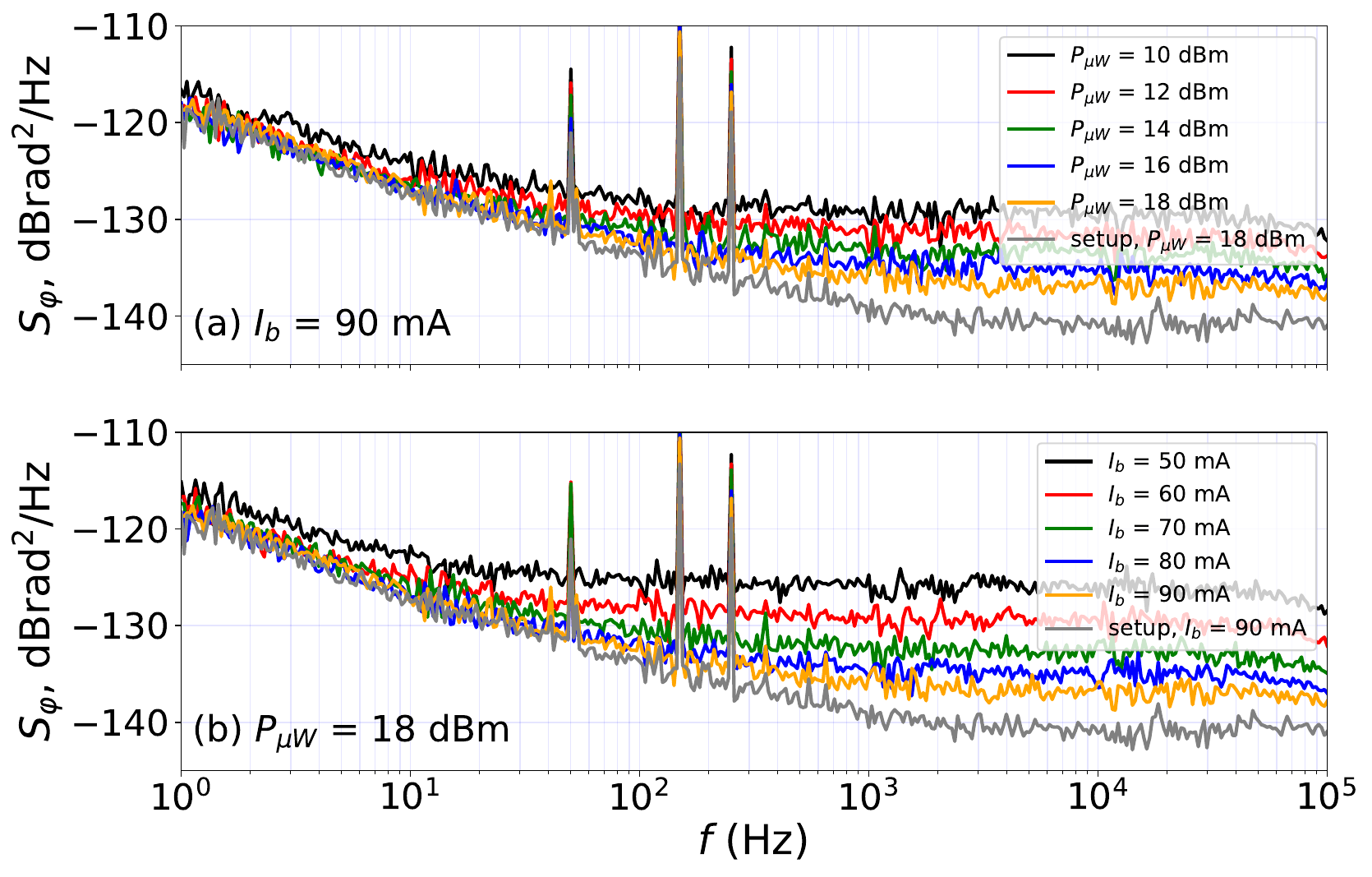}
\caption{Phase noise at 10 GHz with the EOM-based laser system. (a) Fixed bias current ($I_b$ = 90 mA) and different microwave powers $P_{\mu W}$. (b) Fixed microwave power ($P_{\mu W}$~=~18 dBm) and various $I_b$ values. The background noise of the setup is also reported ($P_{\mu W}$ = 18 dBm).  }
\label{figure2}
\end{figure}
\begin{figure}[t]
\centering
\includegraphics[width=\linewidth]{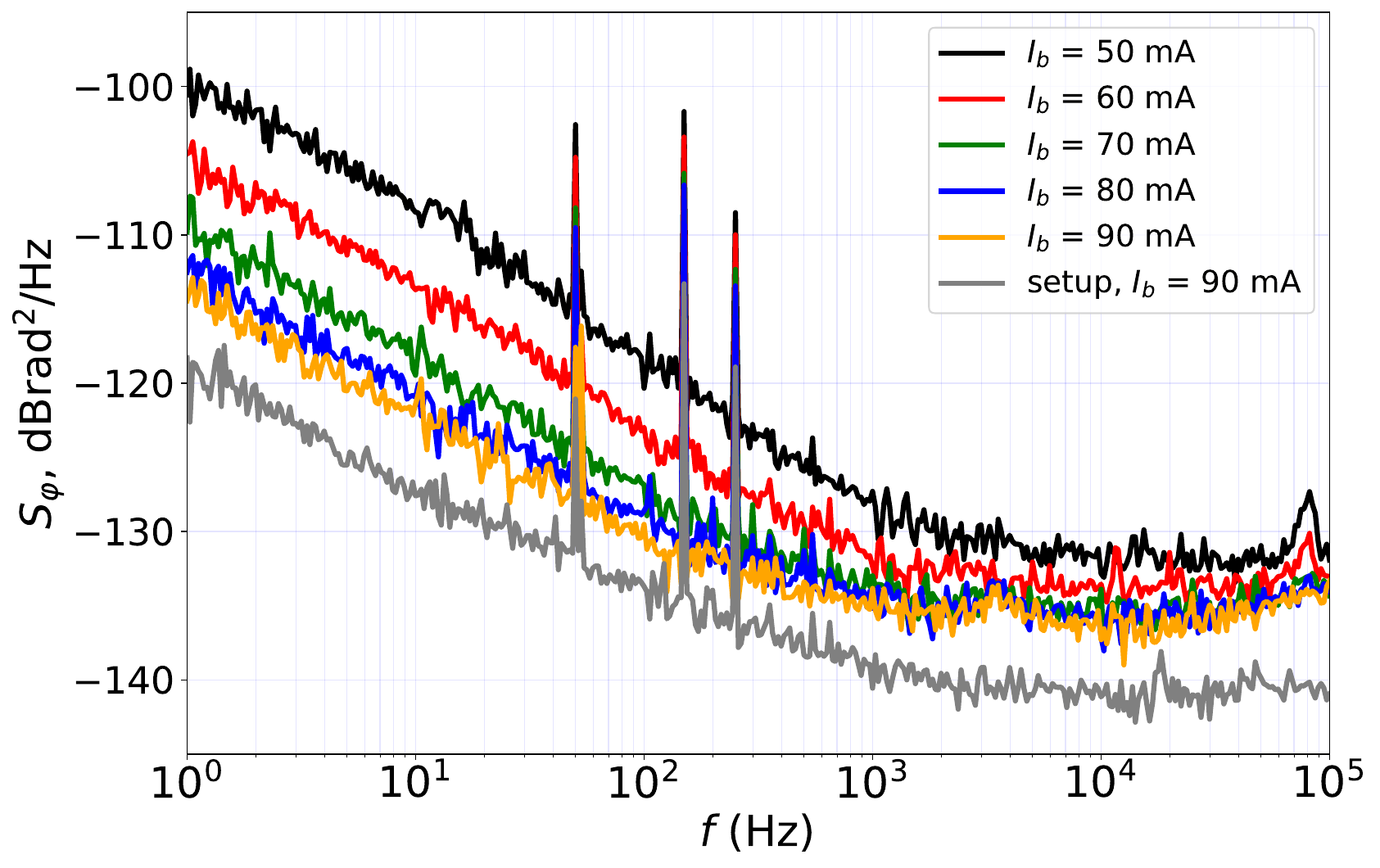}
\caption{Phase noise at 10 GHz with the DM-laser system, for $P_{\mu W}$ = 18 dBm, and different values of $I_b$. The background noise of the setup is also reported ($I_b$ = 90 mA).}
\label{figure3}
\end{figure}

\begin{table}[t]
\caption{Experimental conditions for plots shown in Fig.~2--3.  The quantities $P_{\mu W}$, $P_o$ , $P_{\mu 0}$ and $k_{\varphi}$ are the microwave power at the synthesizer output, the optical power at the photodiode input, the microwave power at the amplifier input, and the mixer phase-to-voltage gain, respectively. }
\label{tab:1}\vspace*{-0.5ex}
\begin{center}\begin{tabular}{|c|c|c|c|}\hline
\multicolumn{4}{|c|}{(A) parameters of Fig. \ref{figure2}(a) }\\
\hline
$P_{\mu W}$ (dBm) & $P_o$ (mW) & $P_{\mu 0}$ (dBm) & $k_{\varphi}$ (mV/rad)\\
\hline
10 & 3.75 & $-$30.65 & 179\\[-1ex]
12 & 3.75 & $-$28.65 & 223\\[-1ex]
14 & 3.75 & $-$26.65 & 268\\[-1ex]
16 & 3.75 & $-$24.7 & 320\\[-1ex]
18 & 3.75 & $-$23 & 371\\
\hline
\multicolumn{4}{c}{~}\\[-2ex]
\hline
\multicolumn{4}{|c|}{(B) parameters of Fig. \ref{figure2}(b) }\\
\hline
$I_b$ (mA) & $P_o$ (mW) & $P_{\mu 0}$ (dBm) & $k_{\varphi}$ (mV/rad) \\
\hline
50 & 1.8 & $-$30.6 & 207\\[-1ex]
60 & 2.3 & $-$27.9 & 258\\[-1ex]
70 & 2.8 & $-$25.8 & 296\\[-1ex]
80 & 3.3 & $-$24.2 & 334\\[-1ex]
90 & 3.75 & $-$23 & 371\\
\hline
\multicolumn{4}{c}{~}\\[-2ex]
\hline
\multicolumn{4}{|c|}{(C) parameters of Fig. \ref{figure3}}\\
\hline
$I_b$ (mA) & $P_o$ (mW) & $P_{\mu 0}$ (dBm) & $k_{\varphi}$ (mV/rad) \\
\hline
50 & 0.45 & $-$30.6 & 200\\[-1ex]
60 & 0.85 & $-$27.9 & 245\\[-1ex]
70 & 1.3 & $-$25.8 & 290\\[-1ex]
80 & 1.95 & $-$24.2 & 323\\[-1ex]
90 & 2.7 & $-$23 & 363\\
\hline
\end{tabular}
\end{center}
\end{table}

\section{Experimental results}
We express the phase noise in terms of power spectral density $S_\varphi(f)$ of the random phase $\varphi(t)$, as a function of the Fourier (modulation) frequency $f$.  Its physical dimension is rad$^2$/Hz.  The usual polynomial approximation of $S_\varphi(f)$ is in our case limited to flicker and white terms, $\mathsf{b}_{-1}/f$ and $\mathsf{b}_0$, respectively.  Their units are $\mathrm{rad^2}$ for $\mathsf{b}_{-1}$, and $\mathrm{rad^2/Hz}$ for $\mathsf{b}_{0}$.  Albeit the quantity $\mathscr{L}(f)$ is more often seen in the literature, defined as $\tfrac{1}{2}S_\varphi(f)$ \cite{IEEE-1139}, $S_\varphi(f)$ should be preferred because it is expressed in SI units, while $\mathscr{L}(f)$ is not. 
We encourage reading Ref.~\cite{Rubiola:Companion} for a tutorial on phase noise and Ref. \cite{IEEE-1139} for the commonly agreed terminology.

Figure \ref{figure2}(a) shows the phase noise $S_{\varphi}(f)$ of the EOM configuration at 10 GHz carrier frequency, with fixed laser bias current ($I_b=90$ mA) and different values of the microwave power $P_{\mu W}$ from the microwave source.  The experimental conditions are reported in Table \ref{tab:1}(A).  The flicker coefficient is $-117~\mathrm{dBrad}^2$, almost independent of the microwave power.  This fact is consistent with the concept of phase modulation from a near-DC $1/f$ fluctuation that we have described in \cite{Boudot:UFFC:2012}.  The value reported is only an upper bound because it equals the background noise of the setup.  Nonetheless, a slight degradation shows up at the lowest power, $P_{\mu W}=10~\mathrm{dBm}$.  More sophisticated methods \cite{Rubiola:RSI:2002} can be envisioned for determining the flicker limitation set by the EOM\@.  However, the level we measured is already lower than the phase noise of state-of-the-art microwave sources \cite{Xie:2016, Grop:EL:2010, Fluhr:2016}.

Still on Fig.~\ref{figure2}(a), we see that the white region of the background noise is $-141~\mathrm{dBrad^2/Hz}$ at 90 mA bias and at maximum microwave power.  This value is satisfactory, to the extent that it is a few dB lower than that of the photonic channel.
The background noise is due to the amplifier noise (thermal noise and noise figure), and to the shot noise.  The former is $S_\text{amp}=FkT=1.25{\times}10^{-20}~\mathrm{W/Hz}$ where $F=3$ (4.8 dB) is the noise factor, and $kT=4{\times}10^{-21}\mathrm{~J}$ is the thermal energy at room temperature.  The latter is $S_\text{sh}=2qIR=2.5{\times}10^{-20}$ W/Hz on the $R=25~\Omega$ load (the shot current sees two 50~$\Omega$ loads in parallel, one inside the diode and one inside the amplifier), but only half of this is transferred to the amplifier.  Thus, amplifier and shot noise give nearly equal contributions, and the overall noise is $N=2.5{\times}10^{-20}~\mathrm{W/Hz}$.  According to \cite{Boudot:UFFC:2012}, the phase noise is $S_\varphi(f)=N/P_{\mu0}=5{\times}10^{-15}~\mathrm{rad^2/Hz}$ ($-143~\mathrm{dBrad^2/Hz}$) with $P_o=5~\mu W$ ($-23~\mathrm{dBm}$).
There is a discrepancy of 2 dB\@.  This is due to a technical problem inside the amplifier.  A separate test on the amplifier alone shows that the white noise exceeds $FkT/P_{\mu0}$ when the input power at one of the four stages is comparable or higher than the compression point, resulting in reduced gain. 

Then, we measured the phase noise for different values of $I_b$, from 50~mA to 90~mA in 10~mA steps.  The results shown in Fig.~\ref{figure2}(b) are obtained in the conditions detailed in Table~\ref{tab:1}(B).  The flicker coefficient is $-117~\mathrm{dBrad^2}$ for all $I_b$, except for a small degradation at 50~mA\@.  A notable degradation is observed, at fixed microwave power when the EOM is operated off the maximum-sensitivity condition by changing the bias voltage $V_b$.  For example, at $V_b$ = 0.6 V (1.3 V below the optimal operation point), the flicker was degraded to $-100~\mathrm{dBrad^2}$.

\begin{figure}[t]
\centering
\includegraphics[width=\linewidth]{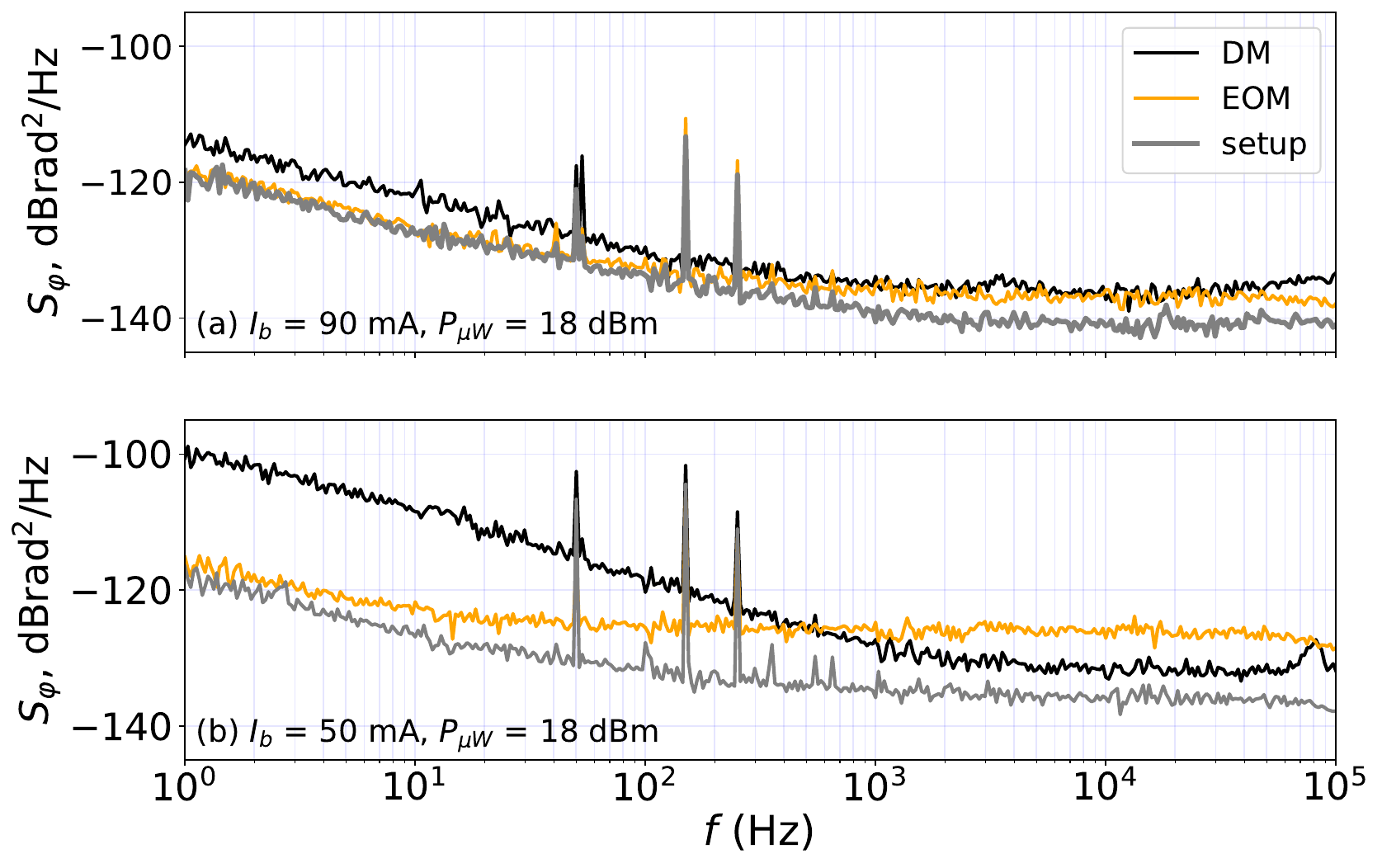}
\caption{Comparison of phase noise spectra obtained for the DM-and EOM-based laser systems. The setup background noise is also reported for information. (a) $I_b$ = 90 mA and $P_{\mu W}$~=~18 dBm. (b) $I_b$~=~50 mA and $P_{\mu W}$~=~18 dBm. }
\label{figure4}
\end{figure}
\begin{figure}[t]
\centering
\includegraphics[width=\linewidth]{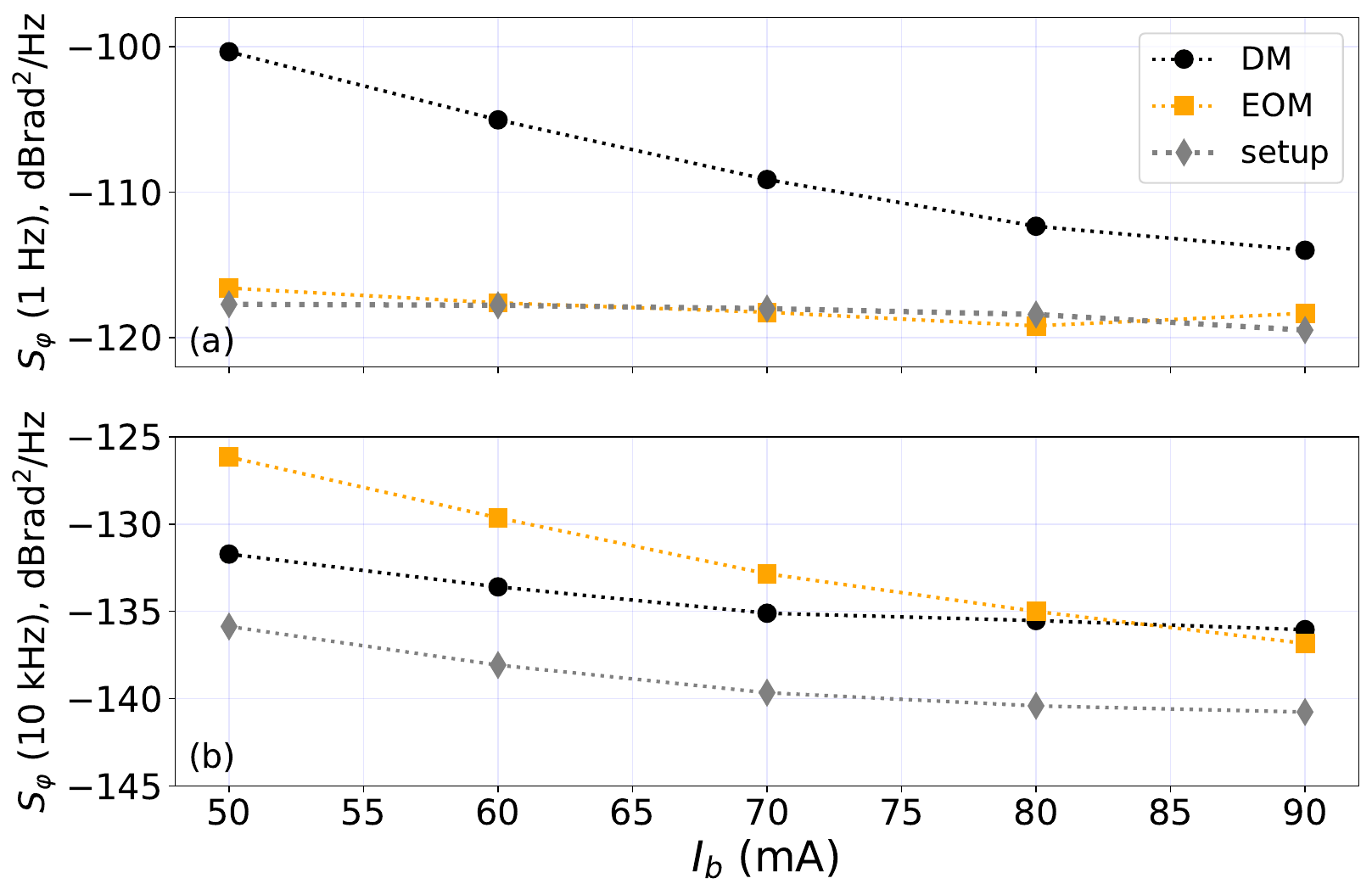}
\caption{Phase noise at 10 GHz at $f$ = 1 Hz (a) and $f$~=~10~ kHz (b) for the DM-laser, the EOM-system and the measurement setup.}
\label{figure5}
\end{figure}

Figure \ref{figure3} shows results obtained with DM of the laser current, at fixed microwave power ($P_{\mu W}$ = 18 dBm) and for several values of $I_b$ from 50 mA to 90 mA in 10 mA steps.  The lowest phase flicker, $-114~\mathrm{dBrad^2}$, is obtained at 90~mA\@.  Reducing the bias current results in a relevant degradation of the flicker, up to $-100~\mathrm{dBrad^2}$ at 50 mA bias. This behavior is due to the increase of the laser low-frequency amplitude and frequency noise occurring at low bias current, up-converted to the microwave frequency \cite{Bdeoui:EMC:2003, Kassa:IEEE:2013, Ahmed:IEEE:2001}. 

Figure \ref{figure4} provides a direct comparison between the two options, EOM and DM\@.  At the highest bias value (Fig. \ref{figure4}~(a), $I_b=90$~mA and $P_{\mu W}=18$~dBm), the DM suffers from slightly higher flicker, while the white noise is nearly equal and just a few dB higher than the background noise of the setup.  In contrast, at low bias (Fig.\ref{figure4}~(a), $I_b=50$~mA and $P_{\mu W}=18$~dBm)
the flicker noise of the DM laser gets 15 dB higher, while the EOM system is not affected.  The white noise of both increases but, surprisingly, the DM features lower white noise.

Finally, Fig.~\ref{figure5} shows the typical phase flicker (phase noise at $f=1$~Hz) and white (phase noise at $f=10$~kHz) for the two options, measured with $P_{\mu W}=18$~dBm.  The flicker noise of microwaves obtained with the DM-laser is higher than with the EOM system for all bias current values. For bias values lower than 80 mA, the noise floor obtained with the EOM is higher than with the DM-laser system and found to increase when $I_b$ decreases.

\section{Conclusions}
We have reported on the phase noise of a photonic microwave channel at 10 GHz, comparing the cases where direct modulation (DM) of the laser current or external electro-optic modulation is used. The laser system using the EOM features a flicker noise coefficient lower than $-117~\mathrm{dBrad}^2$ in a large range of laser bias current, and microwave power.  This is an upper bound because measurement is limited by the background noise of the setup.  Conversely, the white noise floor of the setup is not a limitation, even accounting for the 2-dB excess noise due to the microwave amplifier.  
The flicker noise of the DM scheme is higher than that of the EOM scheme.  We observed $-114$~dBrad$^2$ for a bias current of 90~mA, degraded to $-100$~dBrad$^2$ for a bias current of 50 mA.  
For both configurations, flicker phase noise levels obtained under proper conditions are compliant with the transfer of the most stable microwave signals.

\section*{Funding}
This work has been partly funded by Agence Nationale de la Recherche (ANR) in the frame of the LabeX FIRST-TF (Grant ANR 10-LABX-0048), EquipeX Oscillator IMP (Grant ANR 11-EQPX-0033), and EIPHI Graduate school (Grant ANR-17-EURE-0002) projects.

\section*{Acknowledgments}
The authors thank Vincent Giordano (Institut FEMTO-ST) for fruitful discussions and reading of the manuscript, and Ghaya Baili (Thales RT) and Brian Sinquin (Institut FOTON), for the providing of some bibliography references. 

\section*{Disclosures}
The authors declare no conflicts of interest.

\section*{Data availability statement}
The data that support the findings of this study are available
from the corresponding author upon reasonable request.\\



\end{document}